\documentclass{PoS}

\PoS{PoS(HEP2005)074}
\title{\boldmath Measurement of the $t\bar{t}$ production cross section in $p\bar{p}$ collisions at $\sqrt{s}=1.96$~TeV using multijet events}

\ShortTitle{Top pair production at CDF using multijets events}

\author{\speaker{Lucio Cerrito}\thanks{On behalf of the CDF Collaboration}\\
        University of Oxford, UK\\
        E-mail: \email{lucio@fnal.gov}}

\abstract{We present two measurements of the $t\bar{t}$ production cross section in collisions of protons and antiprotons at $\sqrt{s}=1.96$ TeV. We analyze a dataset of 310$\pm$20 pb$^{-1}$ collected with the CDF$\quad$2 detector. In the first measurement, we select events with six to eight jets, at least one of which having a displaced secondary vertex, little or no missing transverse energy and optimized kinematical criteria consistent with the $t\bar{t}$ all-hadronic decay channel. In the second measurement, we select events with four or more jets, at least one of which having a displaced secondary vertex, high missing transverse energy, and optimized kinematical criteria consistent with the decay of $t\bar{t}$ to $\tau$+jets. The averaged $t\bar{t}$ production cross section, determined from six different measurements using the CDF$\quad$2 detector in the dilepton, lepton+jets and all-hadronic decay channels, is also calculated to be $\sigma_{t\bar{t}}=7.1\pm0.6 ({\rm stat})\pm 0.7(\rm syst)\pm 0.4 ({\rm lumi})\quad{\rm pb}$, in agreement with the prediction of the standard model.}

\FullConference{International Europhysics Conference on High Energy Physics\\
                 July 21st - 27th 2005\\
                 Lisboa, Portugal}

\begin{document}

\section{Introduction}
Top quarks are produced at the Tevatron Collider predominantly in pairs of top-antitop via quark-antiquark scattering ($\sim$85\%) and gluon-gluon fusion ($\sim$15\%). The measurement of the $t\bar{t}$ production cross section provides a test of the theory of Quantum Chromo Dynamics while the comparison of the measurements from different decay channels allows probing both the production and decay mechanisms of the standard model (SM). At $\sqrt{s}=1.96$ TeV, the predicted $t\bar{t}$ cross section is $\sigma_{t\bar{t}}=6.7^{+0.7}_{-0.9}$ pb for the top mass $m_t=175$ GeV/$c^2$ \cite{theory}. The top decays almost exclusively to $Wb$, hence the final state of a $t\bar{t}$ pair can be classified according to the two $W$'s decay, leading to a lepton plus jets, dilepton or all-hadronic final state. The $t\bar{t}$ production cross section has been measured  in all decay channels by the CDF Collaboration during Run I of the Tevatron Collider, using $\sim100\quad{\rm pb}^{-1}$ of data at $\sqrt{s}=1.8$ TeV \cite{run1}. However, the measurements were affected by large statistical uncertainties. 

We present two measurements of the $t\bar{t}$ production cross section at $\sqrt{s}=1.96$ TeV in all-jets final states, using a sample of $310\pm20$ pb$^{-1}$ of $p\bar{p}$ collision data collected with the CDF~2 detector in the years 2002$-$2004. In the first analysis, the event selection isolates a sample of events with 6 to 8 jets and little $\not \!\!\!E_T$ \cite{ft0}, consistent with the hadronic decay of both $W$'s. The second analysis selects a sample with 4 or more jets and large $\not \!\!\!E_T$, consistent with one of the $W$'s decaying to a $\tau+\nu$ and the second hadronically. Finally, we present the average of six $t\bar{t}$ cross section measurements from different decay channels using the CDF 2 detector.

\section{All-jets final state}
\noindent The dataset is collected with a multijet trigger requiring at least four calorimeter clusters with $E_T\geq 15\quad{\rm GeV}$ and total transverse energy exceeding 125~GeV. Events with high missing transverse energy ($\not \!\!\!E_T/\sqrt{\sum E_T}\geq3\quad{\rm GeV}^{1/2}$), and those with a good high transverse-momentum central electron or muon, are rejected in order to avoid that the sample overlaps those used in other top production cross section measurements. The multijet trigger has an estimated 14~nb cross section and is $\sim$85\% efficient in selecting all-hadronic $t\bar{t}$ events. However, the signal-to-background ratio of this preselected sample amounts to about 1/3500, due to the large QCD multijet production of non-top events. An optimized kinematic selection is used to enhance the top signal, and is determined from Monte Carlo simulation \cite{ft1} of signal (S) and background (B) events in such a way as to maximize the ratio $S/\sqrt{S+B}$. The decay of top pairs is characterized by highly energetic, central, spherical events and the optimal selection exploits these kinematic properties by using the number of jets  in the event ($6\leq N_{\rm jets}\leq 8$), high event total energy ($\sum E_T\geq$ 280 GeV), centrality ($C$$\geq$ 0.78) and aplanarity ($A+0.005\sum_3 E_T\geq 0.96$) \cite{ft2}. 

\begin{figure}[t]
\begin{center}
\begin{tabular}{cc}
\includegraphics[width=6.5cm]{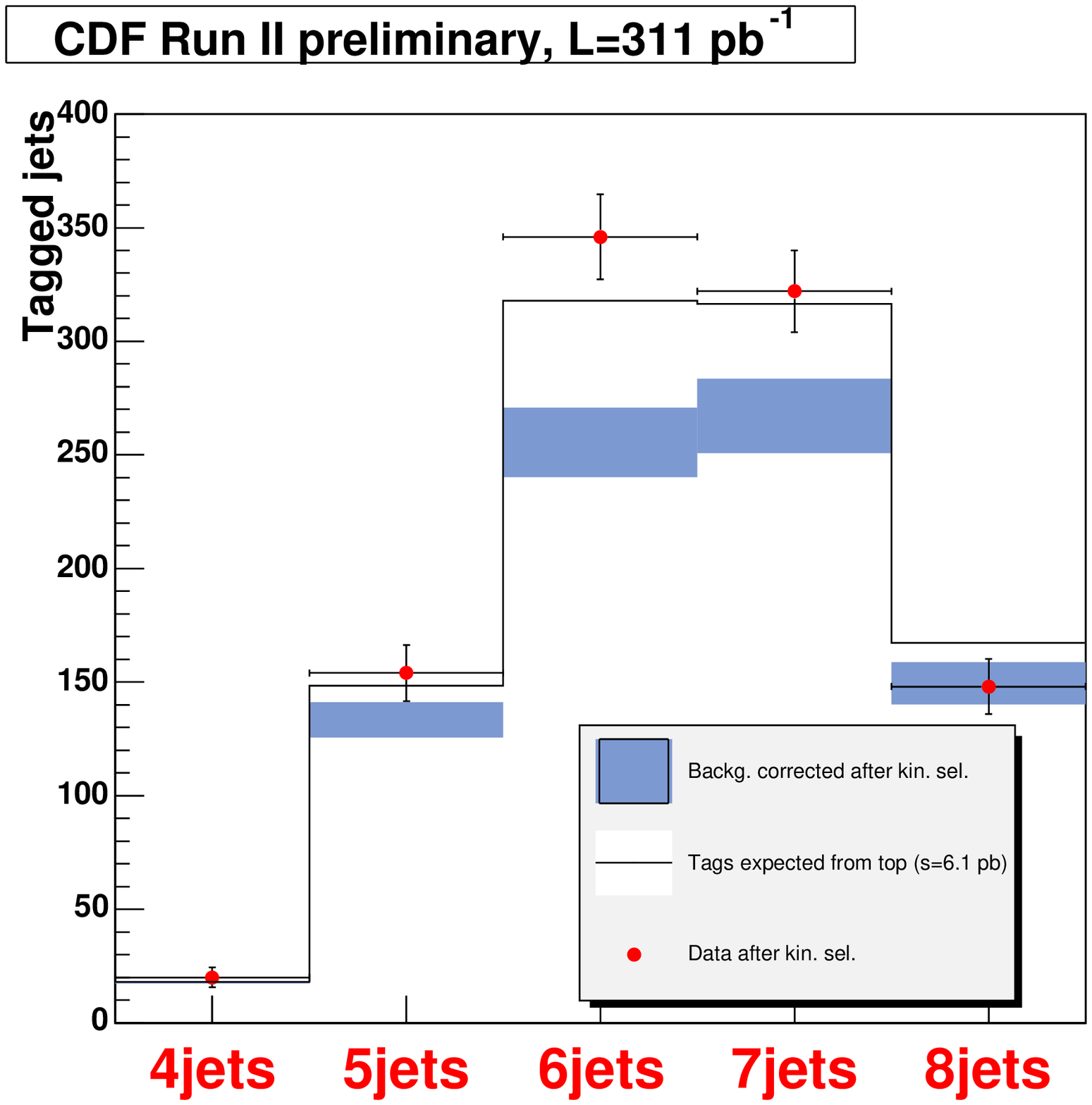} &
\includegraphics[width=6.5cm]{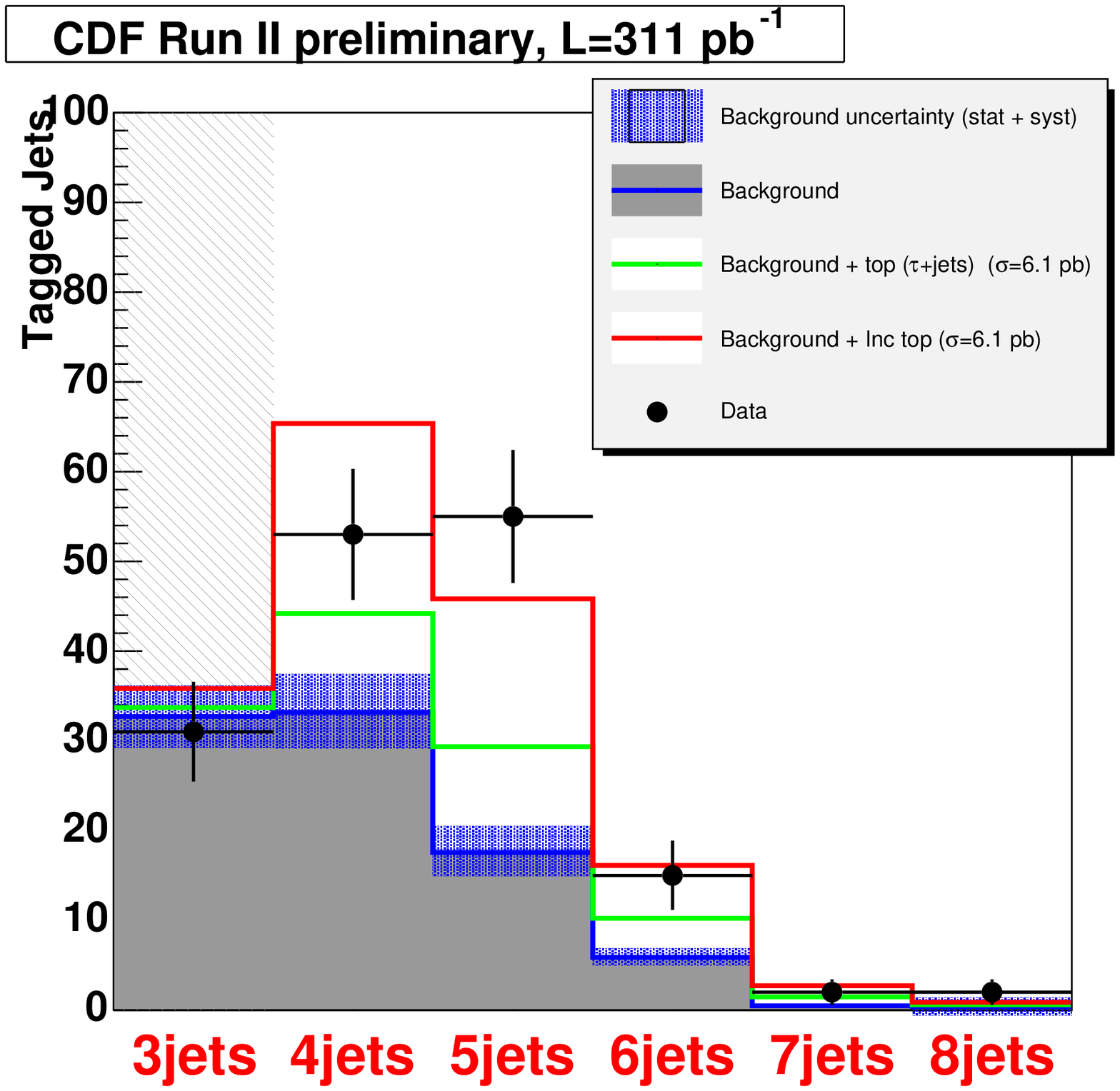} \\
(a) & (b) \\
\end{tabular}
\end{center}
\caption{\small{Comparison between observed and expected tags in (a): the all-jets sample and (b): the~ $\not \!\!\!E_T$+jets sample.}}
\label{fig:allhad}
\end{figure}

With these requirements, the $S$/$B$ ratio is enhanced to about 1/25, with efficiency of accepting $t\bar{t}$ events of $6.6\pm$1.4\%. Finally, events are required to have at least one jet with a displaced secondary vertex (SecVtx tag), compatible with the long lifetime of a $B$ hadron from the top decay. SecVtx tagging enhances $S$/$B$ to about 1/5 with 84$\pm$7\% efficiency of accepting a $t\bar{t}$ event. The selected sample consists of 3342 events with 816 tags. The background is estimated constructing a look-up table of the rate of tags in four-jets events (a highly enriched non-top QCD sample), as a function of the jet-$E_T$, the number of tracks reconstructed in the vertex detector, and the number of reconstructed primary vertices in the event. The expected number of tags due to background in the signal candidate sample is determined using the look-up table and the excess of observed tags is attributed to $t\bar{t}$ events. Figure \ref{fig:allhad}(a) shows the comparison between observed and expected tags as a function of the jet multiplicity in the event. 

From the excess of tags in the signal region, and assuming the top mass ($m_t$) to be 175$\quad$GeV/$c^2$, we determine the $t\bar{t}$ production cross section: $\sigma_{t\bar{t}}=8.0\pm1.7 ({\rm stat})^{+3.3}_{-2.2} ({\rm syst})^{+0.5}_{-0.4} ({\rm lum})\quad{\rm pb}$, where the uncertainties are statistic, systematic and due to the luminosity determination respectively. The systematic uncertainty is predominantly due to the jet energy calibration, contributing $\pm$20\% to the estimated event reconstruction efficiency. Smaller contributions are due to the background estimate ($\pm$10\%), and the Monte Carlo simulation of the $t\bar{t}$ sample.
                                                                                                    
\section{\boldmath $\not \!\!\!E_T$+jets final state}
\noindent The dataset is selected with the same multijet trigger used to isolate the sample of all-jets $t\bar{t}$ events as described above, but the kinematic event selection is replaced by $N_{\rm jets}\geq 4$, large $\not \!\!\!E_T$ ($\not \!\!\!E_T/\sqrt{\sum E_T}\geq4\quad{\rm GeV}^{1/2}$) and $\Delta\phi (\not \!\!\!E_T, {\rm jet})\geq 0.4\quad{\rm rad}$. The high-$\not \!\!\!E_T$ requirement ensures that events are consistent with the presence of a neutrino from the $W$ boson decay. The minimum angular distance between $\not \!\!\!E_T$ and jets rejects preferentially non-top QCD multijet background. Events with a good high transverse-momentum central electron or muon are rejected in order to avoid that the sample overlaps those used in other top production cross section measurements. The S/B ratio after kinematic selection is about 1/5 with the $t\bar{t}$ event selection efficiency of 6.7$\pm$1.4\%. Finally, at least one jet in the event is required to have a SecVtx tag, compatible with the long lifetime of a $B$ hadron from the top decay. SecVtx tagging enhances $S$/$B$ to about 1/1 with 79$\pm$4\% efficiency of accepting a $t\bar{t}$ event. The number of SecVtx tags as a function of jet multiplicity is shown in Figure \ref{fig:allhad}(b). The excess of tags in the four or more jets multiplicity bins is attributed to $t\bar{t}$ signal and assuming $m_t=175\quad{\rm GeV}/c^2$ we measure the cross section: $\sigma_{t\bar{t}}=6.1\pm1.2 ({\rm stat})^{+1.3}_{-0.9}({\rm syst})^{+0.4}_{-0.3} ({\rm lum})\quad{\rm pb}$. 

Systematic uncertainties are for the largest part due to the uncertainty on the efficiency of the multijet trigger (with relative uncertainty $\pm$15\%). Smaller contributions are due to the background estimate ($\pm$10\%), and the Monte Carlo simulation of the $t\bar{t}$ sample.
             
\section{\boldmath Combination of $t\bar{t}$ cross section measurements}
The $t\bar{t}$ production cross section is measured with the CDF 2 detector using the dilepton, lepton plus jets and all-hadronic final states, as well as using several techniques for the same final state. Six of these measurements, including three in the lepton plus jets channel \cite{lj1,lj3}, one in the dilepton channel \cite{dil} and the two measurements presented in this article using the all-jets and $\not \!\!\!E_T$+jets final states, are combined to yield the average CDF $t\bar{t}$ cross section measurement \cite{ft3}. The average is calculated with the BLUE method \cite{blue1,blue2,blue3}, where we account for statistical and systematic correlation, and yields the most precise determination of the $t\bar{t}$ production cross section at $\sqrt{s}=1.96$ TeV to be \cite{combo}: $\sigma_{t\bar{t}}=7.1\pm0.6 ({\rm stat})\pm 0.7(\rm syst)\pm 0.4 ({\rm lumi})\quad{\rm pb}.$

\section{Conclusions}
We have measured the $t\bar{t}$ production cross section in the all-jets and large $\not \!\!\!E_T$+jets final states. Both results are in agreement with the prediction from the standard model. The results are combined with four more measurements using the CDF 2 detector and the average value of the $t\bar{t}$ production cross section at $\sqrt{s}=1.96$ TeV is $7.1\pm0.6 ({\rm stat})\pm 0.7(\rm syst)\pm 0.4 ({\rm lumi})\quad{\rm pb}.$



\begin{thebibliography}{99.}
\bibitem{ft0} The transverse component of the energy imbalance in the calorimeter is called missing transverse energy ($\not \!\!\!E_T$).
\bibitem{ft3} In determining the $t\bar{t}$ event reconstruction efficiency and acceptance in the CDF detector the top quark mass is assumed to be 175 GeV/$c^2$.
\end{thebibliography}
\end{document}